\documentclass[preprint,showpacs,preprintnumbers,amsmath,amssymb]{revtex4}

\usepackage{graphicx}
\usepackage{dcolumn}
\usepackage{bm}

\begin{document}


\title{Quantum key distribution based on frequency-time coding: security and feasibility}

\author{Bing Qi}
\affiliation{Center for Quantum Information and Quantum Control
(CQIQC),
Department of Electrical and Computer Engineering,\\
University of Toronto, Toronto, M5S 3G4, Canada
}%

\date{\today}

\begin{abstract}
We establish a security proof of frequency-time coding quantum key
distribution (FT-QKD) protocol by showing its connection to the
squeezed state quantum key distribution protocol, which has been
proven to be unconditionally secure. We also extend the
prepare-and-measure FT-QKD protocol to an entanglement based FT-QKD
protocol which is more appealing in practice. Furthermore, we
propose a correlated frequency measurement scheme of entangled
photon pair by using time resolving single photon detector.
Simulation results show that the FT-QKD protocol can be implemented
with today's technology.
\end{abstract}

\maketitle

\section{Introduction}

One important practical application of quantum information is
quantum key distribution (QKD), whose unconditional security is
based on the fundamental laws of quantum mechanics \cite{BB84,E91}.
While QKD has been conducted through both optical fiber and free
space, the availability of a worldwide fiber network suggests that
single mode fiber (SMF) could be the best choice as the quantum
channel for practical QKD systems.

Two basic requirements on the quantum channel are low loss and weak
decoherence. While the loss of standard SMF is relatively low, the
decoherence introduced by a long fiber link depends on the coding
scheme. Most practical QKD systems are based on either polarization
coding or phase coding. Unfortunately, these two coding schemes
suffer from polarization and phase instabilities in optical fiber
induced by environmental noise. On the contrary, the frequency-time
coding QKD (FT-QKD) scheme proposed in \cite{Qi06} is intrinsically
insensitive to the polarization and phase fluctuations. This
suggests that the FT-QKD could be a more robust solution in
practice.

In \cite{Qi06}, the security of the FT-QKD protocol was intuitively
interpreted as a result of the energy-time uncertainty relation. In
this paper, we provide a security proof of the FT-QKD protocol by
connecting it to the squeezed state QKD protocol \cite{Ralph00,
Hillery00} whose security against the most general attack has been
proven in \cite{Gottesman01}. This connection is built upon the
observation that the frequency and arrival time of a photon is
connected to its momentum and spatial position. In quantum
mechanics, the commutation relation between position and momentum
operator is the same as that between the two quadratures of an
oscillator. So, mathematically, the FT-QKD is equivalent to the
squeezed state QKD, thus Gottesman-Preskill's security proof in
\cite{Gottesman01} can be applied. We remark that one nice feature
of Gottesman-Preskill's proof is that both the BB84 QKD protocol and
the squeezed state QKD protocol (thus the FT-QKD protocol) are
studied under the same scope. This allows us to apply many important
results developed in the BB84 QKD, such as decoy state idea
\cite{Decoy} and the squash model of threshold single photon
detector (SPD) \cite{Squash_model}, into the FT-QKD protocol.

We remark that single-photon continuous-variable QKD protocols
exploring the spatial freedom of photons have also been investigated
\cite{Spatial_QKD}. However, it could be difficult to implement
those protocols over a long distance.

Section II is a review of the prepare-and-measure FT-QKD protocol
\cite{Qi06} . In Section III, we show that mathematically, the
FT-QKD is equivalent to the squeezed state QKD, therefore
Gottesman-Preskill's security proof \cite{Gottesman01} can be
applied. In Section IV, we propose an entanglement based FT-QKD
protocol which is more appealing in practice. In Section V, we
discuss the feasibility of implementing the FT-QKD with today's
technology. We end this paper with a brief conclusion in Section VI.

\section{A brief review of the prepare-and-measure FT-QKD protocol \cite{Qi06}}

Following the cryptographic convention, the two legitimate users in
QKD are named as Alice and Bob, and the malicious eavesdropper is
named as Eve. In the prepare-and-measure FT-QKD, Alice randomly
chooses to use either ``frequency-basis'' or ``time-basis'' to
encode her random bits. In the frequency-basis, one or more random
bits can be encoded on the central frequency of a single-photon
pulse which has a very small linewidth; In the time-basis, one or
more random bits can be encoded on the time delay (defined
relatively to a synchronization pulse) of a single-photon pulse
which has a very small temporal duration. Upon receiving Alice's
photon, Bob randomly chooses to measure either its frequency or its
arrival time. After the quantum transmission stage, Alice and Bob
compare their bases through a public authenticated channel and they
only keep the results when they happen to use the same basis. Given
the conditional variance of Bob's measurement results is below
certain threshold, they can further generate secure key by
performing error correction and privacy amplification.

A schematic diagram of the prepare-and-measure FT-QKD protocol is
shown in Fig.1. In Fig.1, Alice holds two transform-limited single
photon sources: $S_1$ (for frequency coding) can generate
single-photon pulses with a narrow spectral bandwidth (but a large
temporal duration); $S_2$ (for time coding) can generate
single-photon pulses with a small temporal duration (but a broad
spectral bandwidth). We assume that both $S_1$ and $S_2$ have
Gaussian spectra and their spectral bandwidths are
$\sigma_{\omega1}$ and $\sigma_{\omega2}$, respectively. In the
frequency-basis, Alice encodes her bits by randomly modulating the
central frequency of $S_1$. In the time-basis, Alice encodes her
bits by randomly modulating the time delay of $S_2$. A beam splitter
($BS_A$ in Fig.1) is employed to combine the outputs of $S_1$ and
$S_2$ together. At Bob's side, passively determined by another beam
splitter ($BS_B$ in Fig.1), he can either measure the arrival time
of the incoming photon with a time-resolving SPD, or its frequency
(wavelength) with a dispersive element (such as a dispersive grating
which is shown as DG in Fig.1) followed by a SPD array (SPDA).

\begin{figure}[!t]\center
\resizebox{9cm}{!}{\includegraphics{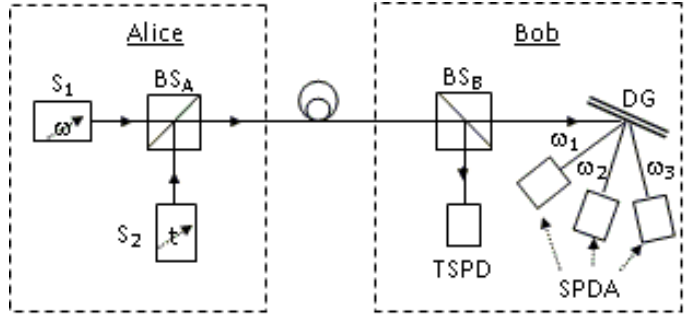}} \caption{Schematic
diagram of the FT-QKD system: $S_1$-narrowband frequency tunable
single photon source; $S_2$-broadband single photon source with
tunable time-delay; $BS_A/BS_B$-beam splitters;
$TSPD$-time-resolving single photon detector; $DG$-dispersive
grating; $SPDA$-single photon detector array.}
\end{figure}

Given the frequency-modulation profile of $S_1$ matching with the
spectrum of $S_2$ and the time-delay-modulation profile of $S_2$
matching with the temporal pulse shape of $S_1$ (as shown in Fig.2),
it can be shown that the density matrix of a frequency-coding photon
is identical to that of a time-coding photon, thus Eve cannot
distinguish them from each other \cite{Qi06}. The security of the
FT-QKD protocol can be intuitively understood from the energy-time
uncertainty relation which puts a constraint on Eve's ability to
simultaneously determine both the frequency and the arrival time of
a photon. Mathematically, Eve's time uncertainty $\Delta_t^{(E)}$
and frequency uncertainty $\Delta_\omega^{(E)}$ (defined as root
mean square (RMS) values) satisfy the following relation
\begin{align}
\Delta_\omega^{(E)}\Delta_t^{(E)}\geq\frac{1}{2}
\end{align}
On the other hand, Bob's measurement uncertainties are not bounded
by equation (1) since he randomly measures either the arrival time
or the frequency of each incoming photon but not both. Alice and Bob
can establish an information advantage over Eve by post-selecting
the cases when they happen to use the same bases, thus secure key
distribution is possible.

\begin{figure}[!t]\center
\resizebox{9cm}{!}{\includegraphics{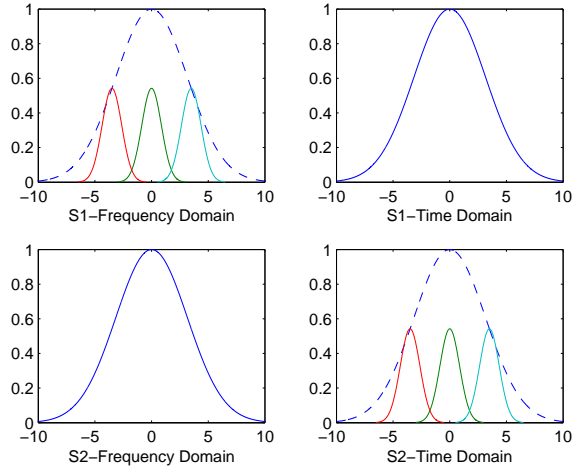}} \caption{Illustration
of the frequency (time) domain modulation profiles of Alice's single
photon sources. Top Left: solid lines-the spectra of $S_1$
corresponding to different frequency shifts; dashed line-the
probabilistic distribution of the frequency-modulation profile which
matches with the spectrum of $S_2$ shown in Bottom Left. Top Right:
the pulse shape of $S_1$ in time domain. Bottom Left: the spectrum
of $S_2$. Bottom Right: solid lines-the pulse shapes of $S_2$
corresponding to different time shifts; dashed line-the
probabilistic distribution of the time-modulation profile which
matches with the pulse shape of $S_1$ shown in Top Right.}
\end{figure}

The FT-QKD protocol can be summarized as follows:

1.Alice generates a binary random number $a$. If $a=0$, she
generates another random number $b$ from the Gaussian distribution
$f_1(b)=(\pi\sigma_{\omega2}^2)^{-1/2}exp[-(b-\omega_0)^{2}/{\sigma_{\omega2}^{2}}]$;
then she sets the central frequency of $S_1$ to $b$ and fires it. If
$a=1$, Alice generates a random number $b$ from the Gaussian
distribution
$f_2(b)=(\pi)^{-1/2}\sigma_{\omega1}exp[-{\sigma_{\omega1}^{2}b^2}]$;
then she sets the time-delay of $S_2$ to $b$ and fires it.

2.Upon receiving Alice's photon, Bob randomly chooses to measure
either its arrival time or its frequency (wavelength).

3.Alice and Bob repeat step 1 and step 2 many times.

4.Through an authenticated classical channel, Alice and Bob
post-select the cases when they use the same bases. After this step,
Alice and Bob share a set of correlated Gaussian variables, which
are called ``key elements''.

5.Alice and Bob convert the``key elements'' into binary bit strings.

6.Alice and Bob can estimate the maximum information acquired by Eve
from the observed quantum bit error rate (QBER). If the QBER is
below certain threshold value, they can future perform error
correction and privacy amplification to distill out a secure key.

\section{Security of the FT-QKD protocol}

In this section, we provide a security proof of the FT-QKD protocol
by connecting it to the squeezed state QKD protocol
\cite{Hillery00}. Section III.A is a brief review of the squeezed
state QKD protocol \cite{Hillery00} and Gottesman-Preskill's
security proof \cite{Gottesman01}. In Section III.B, we apply
Gottesman-Preskill's security proof to the FT-QKD implemented with
perfect single photon sources and ideal photon-number-resolving
SPDs. In Section III.C, we discuss the FT-QKD implemented with weak
coherent sources (attenuated laser source) and practical thresholds
SPDs.

\subsection{The squeezed state QKD and its security proof}

In the prepare-and-measure squeezed state QKD protocol
\cite{Hillery00}, Alice randomly prepares a single mode
electromagnetic field either squeezed at amplitude quadrature
($X_1$) or phase quadrature ($X_2$). For a $X_1$-squeezed state, its
amplitude quadrature is well defined (while its phase quadrature has
a large variance). Alice can randomly modulate the mean value of
amplitude quadrature $\langle X_1\rangle$ to encode her bits;
Similarly, for a $X_2$-squeezed state, its phase quadrature is well
defined (while its amplitude quadrature has a large variance). Alice
can randomly modulate the mean value of phase quadrature $\langle
X_2\rangle$ to encode her bits. At Bob's end, he randomly chooses to
measure either the amplitude quadrature or phase quadrature with a
homodyne detector. After the quantum transmission stage, Alice and
Bob compare their bases for each transmission and only keep the
results when they happen to use the same basis (``key elements'').
If the $\langle X_1\rangle$-modulation profile of the $X_1$-squeezed
state matches with the distribution of $X_1$ of the $X_2$-squeezed
state, and the $\langle X_2\rangle$-modulation profile of the
$X_2$-squeezed state matches with the distribution of $X_2$ of the
$X_1$-squeezed state, then Eve cannot tell which type of squeeze
state Alice has prepared.

Quantum mechanically, the two operators ${X_1}$ and ${X_2}$ are not
commute with each other thus the uncertainty relation applies: $X_1$
and $X_2$ cannot both be defined to arbitrarily high accuracy for a
given quantum state. This is the foundation of the security of the
squeezed state protocol. Its security against the most general
attack is given by Gottesman and Preskill in \cite{Gottesman01}.

Instead of presenting its details, we simply remark that
Gottesman-Preskill's security proof employs quantum error-correcting
codes that encode a finite-dimensional quantum system in the
infinite-dimensional Hilbert space of an oscillator
\cite{Gottesman01_2}. Note, in \cite{Gottesman01}, a pair of
dimensionless position and momentum operators $q$ and $p$ are used
to encode information. The commutation relation between $q$ and $p$
is given by
\begin{align}
[q,p]=i
\end{align}
and the corresponding uncertainty relation is
\begin{align}
\Delta_{q}^{(rms)}\times\Delta_{p}^{(rms)}\geq\frac{1}{2}
\end{align}
where $\Delta_{q}^{(rms)}$ and $\Delta_{p}^{(rms)}$ are defined as
RMS values.

To convert key elements into binary random numbers, the following
distillation protocol is adopted \cite{Gottesman01}: Alice
broadcasts her data modulo $\sqrt{\pi}$, i.e.
$m=mod(S_A,\sqrt{\pi})$; Alice and Bob subtract $m\sqrt{\pi}$ from
their data and correct the remainders to the nearest multiples of
$\sqrt{\pi}$; They extract binary bit values based on whether the
above integers are even or odd.

The secure key rate $R$ of the squeezed state QKD is given by
\cite{Gottesman01}
\begin{equation}
R=\frac{1}{2}[1-f(e)H_{2}(e)-H_{2}(e)].
\end{equation}
Here the factor $1/2$ is due to the fact that half of the time,
Alice and Bob use different bases. $e$ is the observed QBER, $f(x)$
is the bidirectional error correction efficiency, and $H_{2}(x)$ is
the binary entropy function, which is given by
\begin{equation}
H_{2}(x)=-x\log_{2}(x)-(1-x)\log_{2}(1-x).
\end{equation}
Given a perfect error correct code ($f(x)=1$), Equation (4) shows
that as long as the QBER is below $11\%$, secure key distribution is
possible.

The QBER in (4) is determined by \cite{Gottesman01}
\begin{equation}
e\leq\frac{2\Delta}{\pi}exp(-\pi/4\Delta^2)
\end{equation}
where $\Delta^2$ is a measure of the conditional variance of key
elements,
\begin{equation}
Prob(q_A-q_B)=\frac{1}{\sqrt{\pi\Delta^2}}exp[-(q_A-q_B)^2/\Delta^2]
\end{equation}
Here we assume the conditional variance in q-basis ($\Delta_{q}^2$)
is the same as that in p-basis ($\Delta_{p}^2$). In the case of
$\Delta_{q}^2\neq\Delta_{p}^2$, by slightly modifying the protocol
and defining $\Delta^2=\Delta_{q}\Delta_{p}$, (6) is still
applicable \cite{Gottesman01}. In Section III.B, we apply
Gottesman-Preskill's security proof to the FT-QKD protocol.

\subsection{The FT-QKD based on perfect single photon sources and ideal photon-number-resolving SPDs}

We first study the case where Alice holds perfect single photon
sources and Bob has ideal photon number-resolving SPDs.

Note that the energy time uncertainty relation, which has been used
to intuitively understand the security of the FT-QKD protocol, is
fundamentally different from the one applied to a pair of
non-commuting operators, such as $q$ and $p$. This is because in
quantum mechanics, conventionally, time is not treated as an
operator. We remark that there have been great efforts on
establishing a ``time-of-arrival'' operator in quantum mechanics
\cite{time_operator}. Here, instead of touching this deep question
in quantum physics, we simply take an operational interpretation of
the time and frequency measurement. In practice, it is reasonable to
assume that the speed of light in Alice and Bob's station is well
defined and cannot be manipulated by Eve. By choosing a suitable
time reference, the arrival time $t$ and the frequency $\nu$ (or
wavelength $\lambda$) of a single photon are related to its spatial
position $X$ and wave vector $K_X$ (which is proportional to its
momentum $P_X$ as $P_X=\frac{h}{2\pi}K_X$) by
\begin{align}
X=ct/n
\end{align}
\begin{align}
K_X=\frac{2\pi\nu}{c/n}=\frac{2\pi n}{\lambda}
\end{align}
where $n$ is the refractive index, $h$ is the Planck constant and
$c$ is the speed of light in vacuum.

Note that $X$ and $K_X$ satisfy the same commutation relation as $p$
and $q$ do
\begin{align}
[X,K_X]=i
\end{align}
So, mathematically, the FT-QKD is equivalent to the squeezed state
QKD and Gottesman-Preskill's security proof in \cite{Gottesman01}
can be applied.

In practice, there are some technical issues to be resolved. First
of all, in the squeezed state QKD, Bob uses a homodyne detector to
detect Alice's signals. Regardless of the transmission loss, the
homodyne detector always outputs an effective detection result. On
the other hand, SPDs are employed in the FT-QKD. A SPD either
detects nothing or an intact photon. Thus, in the FT-QKD, for each
transmission, Bob has to inform Alice whether he detects a photon or
not. Alice keeps her data only when one of Bob's SPDs clicks. To
take into account of this ``post-selection'' process, we can define
an overall gain $Q_1$ as the ratio of the number of Bob's detection
events to the number of signal pulses sent by Alice. Equation (4) is
then replaced by
\begin{equation}
R=\frac{1}{2}Q_{1}[1-f(e_{1})H_{2}(e_{1})-H_{2}(e_{1})].
\end{equation}
Here, we use subscript $1$ to emphasis the fact that in the FT-QKD,
only single-photon signals contribute to the secure key. Note (11)
is the same as the secure key rate of the BB84 QKD protocol given in
Shor-Preskill's security proof \cite{Shor00}. This is because the
same approach has been adopted in the security proofs of
\cite{Shor00} and \cite{Gottesman01}.

Secondly, in \cite{Gottesman01}, it is assumed that the conditional
variance defined in (7) is solely determined by the squeeze factor
of the source while the contribution of detection system has been
neglected. This is reasonable in the case of the squeezed state QKD
because it is very difficult to prepare highly squeezed state in
practice. However, in the FT-QKD, the uncertainty in quantum state
preparation could be much less than that in the quantum state
detection. For example, both a narrow-band laser pulse with a
spectral linewidth much less than $1pm$ and an ultrashort laser
pulse with a temporal duration less than $1ps$ can be easily
generated in practice. On the other hand, achieving a temporal
resolution better than $10ps$ or a spectral resolution better than
$10pm$ in single photon detection are very challenge. So, in this
paper, we assume that the conditional variance defined in (7) is
determined by the finite temporal and spectral resolutions of the
detection system.

\subsection{The FT-QKD protocol based on weak coherent sources and threshold SPDs}

The security analysis in Section III.B is based on the assumption
that Alice has perfect single photon sources and Bob holds ideal
photon number resolving SPDs. Unfortunately, these ideal devices are
not available yet. A more practical approach is to implement the
FT-QKD with weak coherent sources (heavily attenuated laser sources)
and threshold SPDs (which can distinguish vacuum from non-empty
pulses, but cannot resolve photon number). Fortunately, security
proofs of the BB84 QKD protocol implemented with weak coherent
source and threshold SPD have been developed. As we have remarked
before, one nice feature of Gottesman-Preskill's proof
\cite{Gottesman01} is that both the BB84 QKD protocol and the
squeezed state QKD protocol (thus the FT-QKD protocol) are studied
under the same scope, so many results developed in the BB84 QKD
protocol can be applied to the FT-QKD protocol directly. Instead of
presenting details of previous results, we simply remark that the
decoy state idea \cite{Decoy} can be applied to a FT-QKD protocol
implemented with weak coherent sources and the squash model of SPD
\cite{Squash_model} could be incorporated to resolve the security
issue of using threshold SPDs.

The FT-QKD protocol is interesting in principle. However the system
shown in Fig.1 is too complicated to be attractive in practice. In
Section IV, we extend the FT-QKD to an entanglement based scheme. In
Section V, we discuss the feasibility of implementing the FT-QKD
with today's technology.

\section{The entanglement based FT-QKD protocol}

The prepare-and-measure FT-QKD protocol shown in Fig.1 can be
extended into an entanglement based QKD protocol, as shown in Fig.3.
A source generating energy-time entangled photon pairs can be placed
either at Alice's station or between Alice and Bob. The energy and
time of the two photons in the same pair are Einstein-Podolsky-Rosen
(EPR) \cite{EPR} correlated. One photon from each EPR pair is sent
to Alice and the other one is sent to Bob. Passively determined by a
beam splitter, Alice (Bob) randomly measures either the arrival time
or the frequency (wavelength) of each incoming photon. After the
quantum transmission stage, Alice and Bob compare their measurement
bases for each photon pair and only keep the results when they
happen to use the same basis. The distillation protocol for the
entanglement FT-QKD is the same as the one for the
prepare-and-measure FT-QKD. The entanglement based FT-QKD is closely
related to the squeezed state QKD protocol implemented with two mode
Gaussian entangled squeezed state, whose security has also been
proven in \cite{Gottesman01}.

\begin{figure}[!t]\center
\resizebox{16cm}{!}{\includegraphics{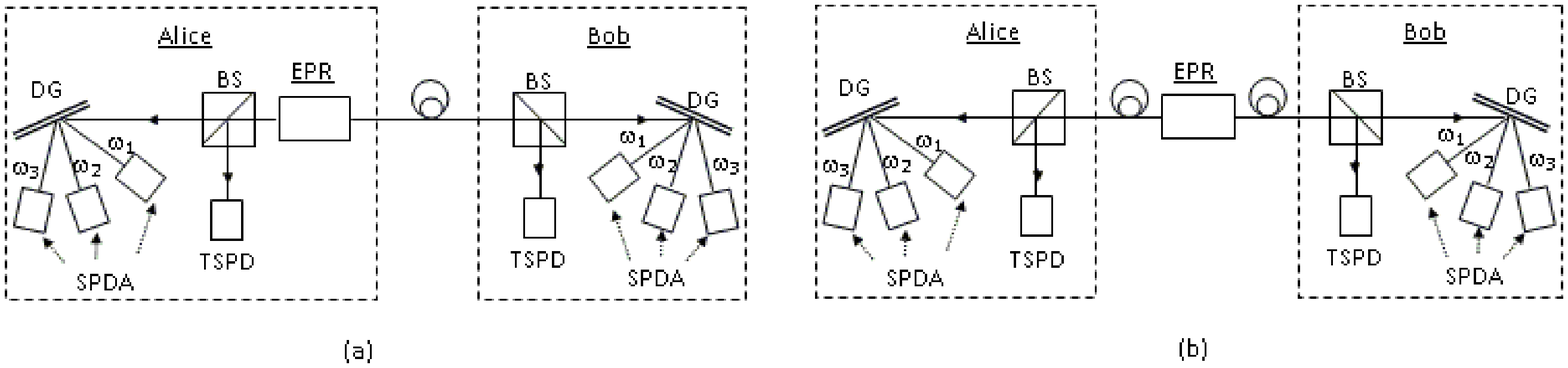}} \caption{Schematic
diagram of the entanglement based FT-QKD system. (a) The EPR source
is placed at Alice's station; (b) The EPR source is placed between
Alice and Bob. EPR-energy-time entangled source; BS-beam splitter;
TSPD-time-resolving single photon detector; DG-dispersive grating;
SPDA-single photon detector array.}
\end{figure}

In practice, the above energy-time entangled photon pairs can be
generated through nonlinear optical processes, such as spontaneous
parametric down-conversion (SPDC). As shown in Fig.4, in this
process, a pump photon spontaneously decays into a pair of daughter
photons in a nonlinear crystal. The conservation of energy and
momentum implies that the generated daughter photons are entangled
in spectral and spatial domains. We assume that the pump pulse has a
narrow spectral bandwidth of $\delta_{\nu P}$ and a relatively large
temporal width of $\delta_{tP}$. We denote the central frequency of
the pump pulse by $\nu_P$. The central frequency of Alice (Bob)'s
photon is $\nu_A(\nu_B)$ and its spectral bandwidth is $\delta_{\nu
A}(\delta_{\nu B})$. Furthermore, Alice (Bob)'s photon is generated
at time $t_A(t_B)$ with a temporal uncertainty of
$\delta_{tA}(\delta_{tB})$.

\begin{figure}[!t]\center
\resizebox{8cm}{!}{\includegraphics{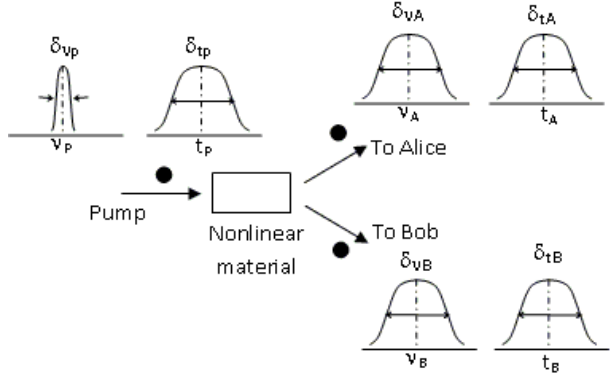}} \caption{Spontaneous
parametric down-conversion (SPDC) process}
\end{figure}

Note the spectral bandwidths $\delta_{\nu A}$ and $\delta_{\nu B}$
of the down-converted photons are determined by the phase matching
condition and the actual experimental setup. The following condition
can be satisfied in practice
\begin{equation}
\delta_{\nu A}\cong\delta_{\nu B}\gg\delta_{\nu P}
\end{equation}
On the other hand, the temporal widths $\delta_{tA}$ and
$\delta_{tB}$ of the down-converted photons are mainly determined by
the temporal width $\delta_{tP}$ of the pump photon
\begin{equation}
\delta_{tA}\cong\delta_{tB}\cong\delta_{tP}
\end{equation}
So each individual photon of an EPR pair has both a broad spectral
bandwidth and a large temporal width, as shown in Fig.4. This
suggests that when Alice and Bob perform time or frequency
measurement, individually, they will observe large uncertainties in
their measurement results. However, if Alice and Bob use the same
basis, their measurement results are highly correlated, i.e.
\begin{equation}
\nu_A+\nu_B\simeq\nu_P
\end{equation}
\begin{equation}
t_A\simeq t_B
\end{equation}

The uncertainty in (14) is determined by the line-width $\delta_{\nu
P}$ of the pump laser, which can be less than $10MHz$. This
corresponds to a wavelength uncertainty in the order of $~0.1pm$ at
telecom wavelength ($\simeq1550nm$). The uncertainty in (15) depends
on the spectral bandwidth $\delta_{\nu A}$ of down-converted photon.
In practice $\delta_{\nu A}$ can be larger than $100GHz$, the
corresponding time uncertainty in (15) is less than $10ps$. For the
detection system, achieving a temporal resolution better than $10ps$
or a spectral resolution better than $10pm$ at single photon level
are very challenge. So we can assume that the conditional variance
defined in (7) is fully determined by the finite temporal and
spectral resolutions of the detection system.

Comparing with the prepare-and-measure FT-QKD protocol based on the
complicated Gaussian modulation scheme, the entanglement based
FT-QKD explores the intrinsic energy-time correlation of an EPR
pair. This greatly simplifies the whole QKD system. Furthermore, in
the FT-QKD system shown in Fig.3, no random numbers are needed
during the quantum transmission stage. This mitigates the
requirement for high speed random number generator \cite{RNG}. The
main technical challenge left is how to achieve high resolution
spectral measurement at single photon level. In Section V, we will
discuss two practical FT-QKD schemes.

\section{Feasibility of the FT-QKD protocol}

\subsection{The prepare-and-measure FD-QKD protocol with discrete modulation}

The FT-QKD protocol shown in Fig.1 can be simplified by using
discrete modulation scheme \cite{Zhu10}, which is shown in Fig.5.

\begin{figure}[!t]\center
\resizebox{8cm}{!}{\includegraphics{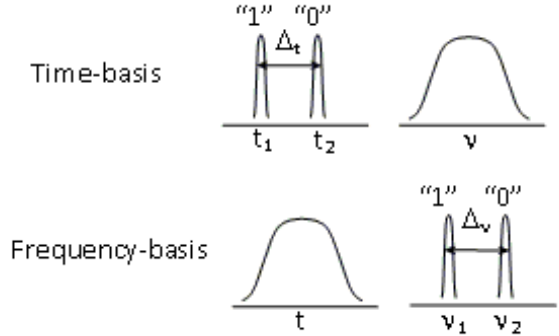}} \caption{The basic
scheme of the FT-QKD protocol with binary modulation}
\end{figure}

In this scheme \cite{Zhu10}, Alice randomly chooses to use either
the frequency-basis or the time-basis to encode her random bit. In
the frequency-basis, Alice uses frequency $\nu_1$ ($\nu_2$) to
encode bit ``1'' (bit ``0''), while in the time-basis, she uses time
delay $t_1$ ($t_2$) to encode bit ``1'' (bit ``0''). At Bob's end,
he randomly measures either the arrival time or the frequency
(wavelength) of each incoming photon. After the quantum transmission
stage, Alice and Bob compare their measurement bases for each
transmission and only keep the results when they happen to use the
same basis. If the QBER is low, they could further generate secure
key by performing error correction and privacy amplification.

Intuitively, to make this protocol secure, Alice's photons in
different bases should at least partially overlap with each other in
both time domain and spectral domain, so Eve cannot distinguish them
faithfully. Furthermore, to apply the energy-time uncertainty
relation to bound Eve's information, the condition of
$(\nu_2-\nu_1)(t_2-t_1)\leq1$ may be required. This put some
constraints on the minimal resolution of Bob's detection system.

From implementation point of view, the FT-QKD with binary modulation
is attractive. However, a security proof for this protocol is still
missing.

\subsection{A practical entanglement based FT-QKD scheme}

Recall that one major technical challenge in the FT-QKD is how to
achieve high resolution spectral measurement at single photon level.
One intuitive idea is to use a highly dispersive element followed by
a time resolving SPD. The dispersive element introduces a
frequency-dependent time delay, thus information encoded in spectral
domain will be transferred into time domain. Thus a time resolving
SPD can be employed to decode the frequency of the incoming photon
by measuring its arrival time. However, this idea cannot be applied
directly. This is because each individual photon has an intrinsic
time uncertainty (for example, in the order of $ns$) which cannot be
distinguished from the frequency-dependent time delay. Fortunately,
the two photons in an EPR pair are entangled in both spectral and
time domain, so the intrinsic time uncertainty of each individual
photon can be canceled out.

As shown in Fig.6, a dispersive element with a dispersion
coefficient of $D_A$ ($D_B$) is placed at Alice (Bob)'s side for
frequency measurement. The dispersion coefficients of the two
dispersive elements are chosen to satisfy $D_B=-D_A$. By using a
suitable time reference, the detection time $T_A$ of Alice's SPD in
frequency-basis is given by
\begin{equation}
T_A=t_A+D_A(\nu_A-\nu_0)
\end{equation}
where $\nu_0$ is the central frequency of the spectral distribution
of down-converted photon from the SPDC source. For the sake of
simplicity, we assume that the two down-converted photons from each
EPR pair have the same spectral distribution, so $\nu_0=\nu_{P0}/2$,
where $\nu_{P0}$ is the central frequency of the pump pulse.

Similarly, Bob's detection time $T_B$ in frequency-basis is given by
\begin{equation}
T_B=t_B+D_B(\nu_B-\nu_0)
\end{equation}
Using equations (14-17) and the facts that $D_B=-D_A$,
$\nu_0=\nu_{P0}/2$, we can see that $T_A$ and $T_B$ are highly
correlated, i.e. $T_A-T_B\cong0$ with a small variance.

We remark that the above frequency correlation measurement scheme is
the same as the one proposed by J. D. Franson in nonlocal
cancellation of dispersion \cite{Franson}.

As a side note, in (16-17) if we choose $D_B=D_A$, then we have
$T_A-T_B=D_A(\nu_A-\nu_B)=D_A(2\nu_A-\nu_P)$. Since $D_A$ and
$\nu_P$ can be treated as constants, this provides a practical way
to measure the spectrum of down-converted photons from a SPDC
source.

\begin{figure}[!t]\center
\resizebox{9cm}{!}{\includegraphics{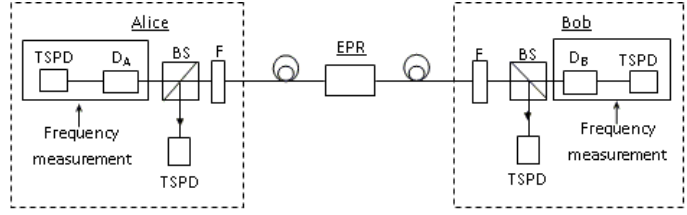}} \caption{Schematic
diagram of a practical entanglement based FT-QKD system:
EPR-frequency-time entangled source; BS-beam splitter; F-spectral
and temporal filters; $D_A$-dispersive component with positive
dispersion coefficient; $D_B$-dispersive component with negative
dispersion coefficient; TSPD-time-resolving single photon detector.}
\end{figure}

In section III, we connected the arrival time and frequency of a
single photon with its spatial position $X$ and wave vector $K_X$.
Similarly, the measurement defined in (16) can be treated as a
 measurement of a combination of $X$ and $K_X$. In general, the above
measurement can be represented by $W=aX+bK_X$, where $a$ and $b$ are
nonzero constants. Since the commutation relation between $X$ and
$W$ is the same as the one between $X$ and $K_X$ (except a scale
factor), Gottesman-Preskill's security proof is still applicable.

Fig.6 is a schematic diagram of the entanglement FT-QKD based on
this new frequency measurement scheme. To evaluate its performance,
equations (6) and (7) can be used to calculate the intrinsic QBER.
As we have discussed above, the conditional variance (thus the
intrinsic QBER) of the FT-QKD is mainly determined by the finite
temporal and spectral resolutions of the detection system.
Specifically, in the entanglement FT-QKD scheme shown in Fig.6,
given the dispersion coefficient of the dispersive elements, the
intrinsic QBER is mainly determined by the time jitter of time
resolving SPDs. For example, commercial dispersion compensation
module based on fiber Bragg grating (FBG) technology can provide a
dispersion coefficient as large as $D_\lambda=7000ps/nm$ with a
moderate loss of $5dB$ \cite{Dispersive}. If the time resolution of
the SPD is $50ps$, then the spectral resolution will be about $7pm$.

In Fig.6, both Alice's and Bob's detection system will make
contributions to the measurement variance. If we assume noises from
Alice and Bob's systems are independent and have identical
distribution ($i.i.d$), then the total variance $\Delta^2$ in (6) is
given by
\begin{equation}
\Delta^2=2\Delta_{X}\Delta_{K}
\end{equation}
where $\Delta_{X}^2$ and $\Delta_{K}^2$ are conditional variances in
time-basis and frequency-basis, respectively.

From (8-9), $\Delta_X$ and $\Delta_K$ are determined by
\begin{equation}
\Delta_X=\frac{c}{n}\Delta_t
\end{equation}
\begin{equation}
\Delta_K=\frac{2\pi n}{\lambda^2}\Delta_\lambda=\frac{2\pi
n}{\lambda^2D_\lambda}\Delta_t
\end{equation}

We remark that variance $\Delta_t^2$ is defined as $1/e^2$, while in
practice, time jitter of SPD ($\delta_t$) is commonly defined in the
fashion of full-width-half-amplitude (FWHA). For a Gaussian
distribution, we have $\Delta_t=\frac{1}{2\sqrt{\ln2}}\delta_t$.
Using (18-20), the conditional variance is given by
\begin{equation}
\Delta^2=\frac{1}{\ln2}\frac{\pi c}{\lambda^2 D_\lambda}\delta_t^2
\end{equation}

Using (6) and (21), we calculate the intrinsic QBER as a function of
the time resolution $\delta_t$ of the SPD. Here, we assume that the
QKD system is operated at telecom wavelength ($\lambda\simeq1550nm$)
and the dispersion coefficient $D_\lambda=7000ps/nm$. The simulation
results are show in Fig.7: the QBER is about $5\%$ for a time jitter
of $70ps$. The time jitter of a state-of-the-art superconducting
nanowire SPD (SNSPD) can be as small as $40ps$ \cite{Dauler10}, and
the resulting QBER is about $0.05\%$.

\begin{figure}[!t]\center
\resizebox{9cm}{!}{\includegraphics{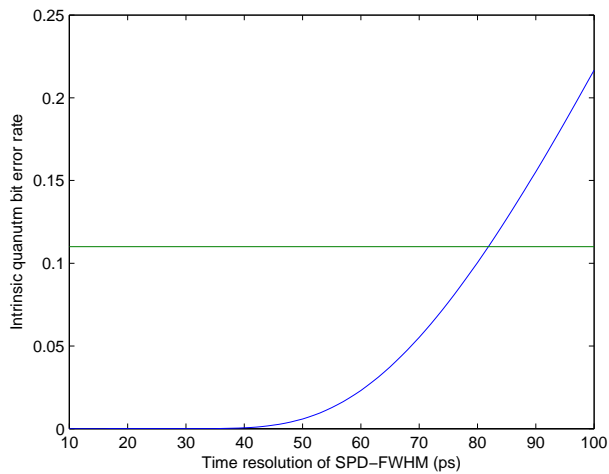}} \caption{Intrinsic
QBER of the entanglement FT-QKD protocol. Here we assume
$\lambda\simeq1550nm$ and $D_\lambda=7000ps/nm$. The QBER is about
$5\%$ for a time jitter of $70ps$. The time jitter of a
state-of-the-art superconducting nanowire single-photon detector
(SNSPD) can be as small as $40ps$ \cite{Dauler10}, and the resulting
intrinsic QBER is about $0.05\%$. We also show the 11\% security
bound in the figure.}
\end{figure}

We remark the secure key rate given by (11) is derived based on
perfect single photon sources (or in the case of entanglement based
protocol, there should be no more than one EPR pair per pump pulse).
However, in practice, multiple pairs could be generated by one pump
pulse, so (11) cannot be applied directly. A similar problem has
been studied in the entanglement based BB84 QKD protocol
\cite{Ma07}, where the SPDC source has been identified as a
basis-independent source thus the security analysis given in
\cite{Koashi03} can be applied. To apply the result in \cite{Ma07}
to the entanglement based FT-QKD system, an appropriate squash model
of the threshold detector is needed.

In practice, the effective detection window of the QKD system may be
limited: in the time-basis, photons arriving outside of a certain
time window may be treated as noise photons and be discarded; in the
frequency-basis, photons outside of certain spectral range may not
be detected. Eve may take advantage of this imperfection and
introduce basis-dependent detection efficiency. Similar security
issues have been studied in the BB84 QKD by, for example, time-shift
attack \cite{time_shift}. To close this potential loophole, spectral
and temporal filters (represented by $F$ in Fig.6) can be placed at
the entrance of the QKD system to make sure that the incoming
photons are within the desired spectral and temporal range.

\section{Conclusion}

One major advantage of the FT-QKD protocol is its robust against
environmental noise:  the frequency/time coding scheme is
intrinsically insensitive to the polarization and phase
fluctuations. This could improve the stability of a practical QKD
system dramatically. One may worry about the temporal broadening of
a narrow laser pulse due to fiber dispersion. Fortunately, the
dispersion of SMF at telecom wavelength has been thoroughly studied
and various dispersion compensation technologies are available. For
example, in \cite{Jiang05}, after passing through a $50km$ fiber, a
$460fs$ pulse was only slightly broaden to $470fs$. This is orders
lower than the time resolution of today's SPD.

In this paper, we establish a security proof of the FT-QKD protocol
by showing its connection to the squeezed state QKD. We also extend
the prepare-and-measure FT-QKD protocol to an entanglement based
FT-QKD protocol which is more appealing in practice. Furthermore, we
propose a correlated frequency measurement scheme by using time
resolving SPD. Simulation results show the feasibility of the FT-QKD
protocol.

As for future research directions, a rigorous security proof of the
FT-QKD based on binary modulation scheme is highly desired. For the
entanglement based FT-QKD, a suitable squash model for threshold
SPDs is required. Furthermore, to fully take advantage of the
continuous variable FT-QKD, a distillation protocol which can
generate more than one bit from each transmission should be
developed.

\textbf{Acknowledgement:} The author is very grateful to Hoi-Kwong
Lo and Li Qian for their support and helpful comments. The author
also thanks John Sipe, Eric Chitambar, Christian Weedbrook, Wolfram
Helwig, Wei Cui, Luke Helt, and Sergei Zhukovsky for helpful
discussions. Financial support from CFI, CIPI, the CRC program,
CIFAR, MITACS, NSERC, OIT, and QuantumWorks is gratefully
acknowledged.

\end{document}